\documentclass{article}

\usepackage{arxiv}
\usepackage{cite}
\usepackage{amsmath,amssymb,amsfonts}
\usepackage{algorithmic}
\usepackage{graphicx}
\usepackage{textcomp}
\usepackage{xcolor}
\usepackage{comment}
\usepackage{url}
\usepackage{booktabs}
\usepackage{algorithmic}
\usepackage{algorithm}
\usepackage[utf8]{inputenc} 
\usepackage[T1]{fontenc}    
\usepackage{hyperref}       
\usepackage{url}            
\usepackage{booktabs}       
\usepackage{amsfonts}       
\usepackage{nicefrac}       
\usepackage{microtype}      
\usepackage{lipsum}
\usepackage{graphicx}
\graphicspath{ {./images/} }

\title{Intelligent Proactive Fault Tolerance at the Edge through Resource Usage Prediction}

\author{
 Theodoros Theodoropoulos \\
  Department of Informatics \& Telematics\\
  Harokopio University of Athens\\
  Athens, Greece\\
  \texttt{ttheod@hua.gr} \\
   \And
 John Violos\\
  School of Electrical and Computer Engineering\\
  National Technical University of Athens\\
  Athens, Greece \\
  \texttt{violos@mail.ntua.gr} \\
  \And
 Stylianos Tsanakas\\
  School of Electrical and Computer Engineering\\
  National Technical University of Athens\\
  Athens, Greece \\
  \texttt{el09727@mail.ntua.gr} \\
  \And
 Aris Leivadeas\\
  Department of Software and IT Engineering\\
  Ecole de technologie superieure\\
  Montreal, Canada\\
  \texttt{aris.leivadeas@etsmtl.ca} \\
  \And
Konstantinos Tserpes\\
  Department of Informatics \& Telematics\\
  Harokopio University of Athens\\
  Athens, Greece\\
  \texttt{tserpes@hua.gr} \\
   \And
Theodora Varvarigou\\
  School of Electrical and Computer Engineering\\
  National Technical University of Athens\\
  Athens, Greece \\
  \texttt{dora@telecom.ntua.gr} 
}

\begin{document}
\maketitle
\begin{abstract}
The proliferation of demanding applications and edge computing establishes the need for an efficient management of the underlying computing infrastructures, urging the providers to rethink their operational methods. In this paper, we propose an Intelligent Proactive Fault Tolerance (IPFT) method that leverages the edge resource usage predictions through Recurrent Neural Networks (RNN). More specifically, we focus on the process-faults, which are related with the inability of the infrastructure to provide Quality of Service (QoS) in acceptable ranges due to the lack of processing power. In order to tackle this challenge we propose a composite deep learning architecture that predicts the resource usage metrics of the edge nodes and triggers proactive node replications and task migration. Taking also into consideration that the edge computing infrastructure is also highly dynamic and heterogeneous, we propose an innovative Hybrid Bayesian Evolution Strategy (HBES) algorithm for automated adaptation of the resource usage models. The proposed resource usage prediction mechanism has been experimentally evaluated and compared with other state of the art methods with significant improvements in terms of Root Mean Squared Error (RMSE) and Mean Absolute Error (MAE). Additionally, the IPFT mechanism that leverages the resource usage predictions has been evaluated in an extensive simulation in CloudSim Plus and the results show significant improvement compared to the reactive fault tolerance method in terms of reliability and maintainability.
\end{abstract}


\section{Introduction} 
During the last decade, the scientific community witnessed the emergence of applications that are intertwined with a set of demanding  QoS requirements. Extended Reality (XR) \cite{makris2021cloud} applications is one instance of this type of application. XR applications are associated with various QoS requirements \cite{theodoropoulos2022cloud} that are based on the ability to provide an immersive end-user experience. These requirements may include aspects such as latency and bandwidth. Studies have shown that for an end-user experience to be acceptable in terms of immersion, the end-to-end latency shall not surpass the 15ms mark, and the bandwidth should be able to reach up to 30 Gbps \cite{boos_demo_2016}. On top of that, the desired integrity of the aforementioned immersive experiences may be jeopardized by faults in task processing, due to potential disruptions in service delivery. Therefore, it is of paramount importance for this class of applications to be able to exhibit fault tolerance capabilities. Furthermore, this class of applications present various demanding requirements in terms of computational resources, since they incorporate the rendering of 3D models and detailed graphics. Because of these computational requirements, monolithic development architectures would result in prohibitively bulky and expensive end-user equipment in order to facilitate the required  computational resources \cite{taleb2022towards}.

\textcolor{black}{Cloud computing is able to partially alleviate the burden that is imposed on the end-user devices by providing computational resources that these applications may run on via the Internet. Cloud computing is based on the use of shared computational resources that may span over multiple locations. Therefore, part of the computational burden is transferred to these shared resources. Unfortunately, the distance between the end-user devices and the cloud servers may result in high latency and low available bandwidth. Thus, the need to bring processing and data closer to the devices where it’s being generated \cite{campolo_towards_2020} was created. In the case of XR applications, these devices may include smart objects, mobile phones, network gateways, sensors and a plethora of immersion devices. This distributed computing paradigm, defined as edge computing aims to establish decentralized topologies and allow the relocation of various computational and storage resources closer to the edge of the network. By doing so, it is expected to provide service delivery and content caching in better response times and transfer rates. The aforementioned devices may vary wildly in terms of computational prowess. As a result, it is necessary to make sure that the computations that take place at the edge are not demanding and do not exceed the computational capabilities of the involved devices.}

When contemplating the nature and requirements of modern-day applications it is of major importance for the workload execution to be resilient and meet the QoS standards set by the industry. The devices at the edge of the network are subjected to significant fluctuations in the amount of offloaded tasks over time  \cite{hameed_machine_2021}. Hence, it is of paramount importance that these fluctuations do not affect the performance of the system and cause process faults \cite{petrosino_dynamic_2020}. In addition to that, edge computing environments are characterized by extreme heterogeneity and dynamicity in regards to the tasks and the processing nodes involved. This unprecedented situation gave birth to the need for an IPFT method which should be robust to infrastructure and workload changes.

Monitoring and predicting the capacity under which the edge nodes are operating in terms of resource metrics such as CPU, RAM, bandwidth and disk can be a valuable piece of information with regards to implementing fault tolerance policies. Resource metrics have high serial and cross correlation values making the use of time series methods rational \cite{nisar_resource_2020}. Regression-based RNN \cite{shiva_prakash_survey_2019}, which leverage time series characteristics through Gated Recurrent Units (GRU) \cite{shen_deep_2018} or Long Short-Term Memory (LSTM) \cite{violos_lstm}, can be used in order to accurately predict the resource metrics. 

In order to handle the extreme heterogeneity and dynamicity of the edge environments, we provide a systematic methodology for building deep learning models in an automatic way using historical data. Common approaches that are based on manual trial and error methodologies in regards to creating acceptable deep learning architectures require many working hours to be spent by deep learning experts every time the deployed applications, user behaviour or the edge infrastructure change. On the other hand, the available deep learning automation methods still have significant shortcomings such as low efficiency and high computational requirements  \cite{yu_hyper-parameter_2020}. A potential solution to these drawbacks could be the extension of evolutionary algorithms, as well as their combination with other models for hypertuning. 

The facts mentioned above motivated us to propose an IPFT method that focuses on processing faults; faults related with resource shortage and the resulting incompetence in regards to processing capabilities that impede the underlying infrastructure to execute tasks within acceptable QoS ranges. Our research goals are to propose a composite deep learning architecture suitable for predicting in a unified way the ability of the edge nodes to execute the incoming workload and an appropriate operational pipeline that guarantees advanced fault tolerance. The cornerstone of this pipeline is the ability to operate in a proactive manner. Whenever a bottleneck in task execution is expected to occur, proactive measures like task migration and node replication should be triggered. A composite deep learning architecture should leverage the time series characteristics of the edge resources and the involved tasks which can be provided by monitoring systems (i.e. Prometheus) \cite{an_pre-study_2021}. Finally, we propose the HBES optimization algorithm in order to provide a composite deep learning model which is nearly optimal. 

The four major contributions of our research are:
\begin{itemize}

\item \textcolor{black}{The proposal of the IPFT method that achieves high reliability and maintainability with very good performance in terms of timely fault detection and repair.}

\item \textcolor{black}{ A discussion of how a specific category of faults, the process faults are related and can be predicted by the resource utilization metrics of processing edge nodes
}
\item The proposal and analysis of the theoretical principles  
of a composite deep learning model for edge resource usage prediction that includes two channels. One with feed-forward and one based on RNN layers.

\item The proposal of an innovative hybrid hyperparameter optimization model that combines the evolution strategy with the Bayesian optimization algorithms in order to gauge a close to optimal composite deep learning architecture.

\end{itemize}

The rest of the paper is structured as follows: Section \ref{Sec:Related} highlights the related work in fault tolerance, resource usage prediction, time series, deep learning and hyperparameter optimisation techniques. Section III explains how a proactive fault tolerance mechanism can leverage resource usage predictions. Section \ref{Sec:RNN} provides an analysis of the RNN multi-output regression approaches, the composite deep learning architectures and the HBES method. Section \ref{Sec:Experimental} describes the experimental setup in a real edge computing dataset, the simulation of IPFT in CloudSim Plus and the evaluation results. \textcolor{black}{Section} \ref{Sec:Conclusion} concludes the paper, reports the current limitations and suggests future directions.

\section{Related Work}
\label{Sec:Related}

Fault tolerance mechanisms are mainly divided in two categories; reactive and proactive. The reactive approach decreases the influence of failures in the edge infrastructure after a failure has actually occurred. The main reactive fault tolerance methodologies are the reactive replication, resubmission, retry and checkpointing. \textcolor{black}{For instance, a state of the art replication-based fault tolerance mechanism in large-scale graph-parallel systems was proposed in \cite{chen_replication-based_2018}, which works supporting cheap maintenance of vertex states. This mechanism replicates the vertices with normal message exchanges, and provides fast in-memory reconstruction of the failed vertices from replicas in other machines. The replication increases the reliability of the system and the chance that the task will finish correctly, at the expense of additional resources for redundancy. }

A retry approach that uses idempotent HTTP methods has been proposed in \cite{noauthor_transient_nodate} for offloading and execution failures. This retry strategy has the advantage that it utilizes the least resources of the computing environment and minimizes the user time, but at the expense of increasing the response time, since HTTP methods may be retried multiple times until they complete successfully. In terms of checkpointing, a reactive fault tolerance approach for the serverless paradigm was investigated in \cite{karhula_checkpointing_2019}. Specifically, the authors through checkpointing and live container migration have succeed to save resources in constrained devices. Unfortunately, also in this case, the execution time is increased since it includes the recovery time of the failed servers.

In the proactive fault tolerance approach a potential fault is predicted in order to avoid its influence on the task execution. Tian et al.  \cite{tian_cloud_2020} use the tree based model which is a statistical analysis technique to diagnose the high risk cloud tasks and apply virtual machine migration techniques. This approach even if it significantly improves the reliability and efficiency, it has generalization limitations since it can not automatically adapt to new computing environments. Machine learning and online learning methods have also been used in combination with microservices architecture and IoT systems to detect fault patterns and pre-emptively mitigate the faults \cite{power_microservices_2018}. Another advanced proactively model has recently been proposed in \cite{Theodoropoulos2022-ys}. This model performs multi-step predictions in order to estimate the process faults and the QoS degradation in different time granularities. This approach utilizes an encoder-decoder model and gauges the ability of the infrastructure to process the incoming tasks at different production rates.

From the above it is evident that the limited computational capacity of processing nodes set barriers to the edge computing and IoT applications \cite{firdose_2021}. We can overcome these barriers through efficient resource management \cite{Dechouniotis_2020}. This process includes the guarantee of well-defined QoS metrics and an accurate workload prediction \cite{masdari_survey_2020}. Some of the workload prediction models leverage RNN and specifically LSTM, formulating the resource usage metrics as data sequences \cite{violos2022intelligent}. None of these approaches explores how a proactive fault tolerance method can leverage the resource usage predictions, but instead choose to focus on the actual resource usage prediction process \cite{Theodoropoulos2023}.

For instance, the authors in\cite{nisar_resource_2020} have used an Autoregressive Integrated Moving Average (ARIMA) to avoid resource under-provisioning or over-provisioning in data centers. ARIMA has the limitation that it models linear dependencies and it is based on the stationarity assumption. However, as noticed in \cite{theodoropoulos_encoder-decoder_2021} the workload to be processed at the edge has trend, seasonality and nonlinearities in the execution behaviours,  which limit the application of statistical linear models. These limitations are overcome by  machine learning models such as K-Means, Decision tree and K-Nearest neighbors \cite{prathibha_investigating_2019}. While there is a lot of classical machine learning models publicly available, for our experiments we selected to use XGBoost \cite{chen_xgboost_2016} because it is popular for winning Kaggle and other prestigious machine learning competitions \cite{nielsen_tree_2016}. XGBoost mostly uses gradient boosted decision trees and is available as an open-source software library.

The limitation of the machine learning approaches is that every time the edge-cloud infrastructure, the user behaviour, or the application change, new models should be trained from scratch with human assistance. Automated machine learning achieves with an automatic way to guide the learning process of models, maximizing the performance, and minimizing the computational budget without human involvement. In the domain of cloud computing, the Application and User Context Resource Predictor (AUCROP) \cite{violos_leveraging_2019} has been proposed for automated usage of classical machine learning algorithms. In addition, a general-purpose automated machine learning meta-model for data pre-processing, regression and hyperparameter tuning through the Bayesian optimization is the Auto-sklearn \cite{feurer_auto-sklearn_2019}. 

Keras-Tuner \cite{bursztein_cutting_2019} is the approach from Keras to automate the hyperparameter tuning, also named hypertuning. Keras is one of the most popular frameworks in the deep learning community. Keras-Tuner has the advantage that the hypermodels, the range of hyperparameters and the tuning process is smoothly integrated in Keras but it supports only the optimizers: (a) random search, (b) hyperband which is a random search with early stopping, and the (c) Bayesian optimization. In our experimental evaluation we used Keras-Tuner, AUCROP and Auto-sklearn. In this work we also extend the research in the hypertuning combining evolution strategy with the Bayesian optimization. Thus, we propose the innovative HBES method as a prominent automated deep learning solution that tackles the heterogeneity and the dynamicity of edge computing environments.

More specifically, our work aims to extend the resource usage prediction method by proposing a proactive fault tolerance method that leverages the resource usage predictions and tackles the above mentioned limitations as follows: Firstly, the IPFT mechanism requires a minimum number of replicas of the execution nodes since it requests a replication only after a fault prediction. Secondly, the IFPT does not include the time overhead of task rescheduling after a fault since the replication and rescheduling of the task will take place timely and proactively. Thirdly, the IPFT leverages deep learning RNN models in order to overcome the limitations of statistical models and adapts to non-linear and non-stationary resource metrics. Lastly, \textcolor{black}{the introduced HBES} qualifies the generality of the whole process which is a common limitation of many methods in the pertinent literature. 

\section{Leveraging Resource Usage Predictions for Proactive Fault Tolerance}

\subsection{Resource Usage Prediction in Edge Computing }
\label{Sec:Predict}
The management and orchestration of edge computing infrastructures can be improved by leveraging various resource utilization metrics. The most notable of these metrics are CPU, RAM, bandwidth, and disk I/O. At the same time, the edge computing paradigm is characterized by the dynamic behaviour and the heterogeneity of the processing edge nodes, which are obliged to operate within some specific constraints dictated by the QoS requirements. 

Hence, the decision making in a dynamic and heterogeneous environment is a rather complex process, which requires every available source of information to be used. Prediction of the resource consumption metrics, by leveraging time series characteristics of historical data, constitutes one of the most valuable piece of information. It serves as a strong indicator for the availability of the processing nodes in order to receive additional workload or to predict potential QoS degradation in future time steps. Accordingly, the publicly available monitoring tools like Prometheus, OpenTSDB, Nagios and InfuxDB can provide the resource metrics in a stream format or in a time series database like PromQL. These time series databases can be used to produce datasets which are suitable for the RNN model training.

The dynamic behaviour of edge nodes is attributed to the fluctuation of application requests and their workload. The number of requests per time interval changes within various time-frames and is affected by many periodic phenomena. Furthermore, the edge is characterized by a high heterogeneity, since  the edge nodes can have different hardware and software characteristics, such as memory, computing power, etc. This heterogeneity becomes more apparent when taking into consideration the various flavours of Raspberry Pis, Arduinos, sensor motes, and other micro-controllers that coexist and collaborate within the same edge infrastructure. 

At the same time, application owners can set strict performance requirements for the edge nodes in terms of availability, throughput and different types of potential delays. Thus, edge providers are struggling to get the QoS metrics within the acceptable ranges specified. Consequently, in order to guarantee that the infrastructure will have sufficient computational capacities to handle fluctuating demands, a fault tolerance mechanism should proactively take decisions considering the amount of resources and the availability of the processing nodes.

In the scientific literature there are different categories of faults that correspond to specific fault tolerance mechanisms. The major categories of the faults are: (a) Network faults, (b) Physical faults, (c) Process faults and (d) Service expiry faults \cite{kumari_survey_2021}. Among these faults, process faults can be very severe. In more detail, process faults occur in processes because of resource shortages which lead to longer task execution delays or even execution stoppages. This type of faults can eventually lead to performance degradation that is not acceptable for real-time and/or time-sensitive applications. 

Thus, it becomes evident that the execution of the tasks has a solid impact on resource utilization and vice versa. This fact has led us to examine the resource utilization metrics in order to take proactive fault tolerance decisions to reduce the adverse effects of process faults.

\subsection{Proactive Fault Tolerance}
Given that the modus operandi of the edge computing paradigm relies on a vast number of compute nodes operating simultaneously, it is extremely important to consider component failures as an inevitability. By doing so, it is important to ensure that the infrastructure will continue operating without interruptions and QoS deterioration. The main way of ensuring this service continuity, is by triggering migration policies and by utilizing backup components, which automatically replace the failed ones in a manner which guarantees the QoS. 

The replication process of a node, such as a virtual machine, requires a certain waiting period, which would provoke QoS degradation. Thus, fault tolerance should be achieved by following a proactive approach. At any given time, the network should contain a specific number of computational nodes, which can remain idle until one of the already working components ceases to function properly. Given that redundant computational nodes may be requested, it is important to keep this redundancy to a minimum. However, by utilizing machine learning algorithms, it is possible to extract information regarding the behaviour patterns of the services and the process faults that occur. Hence, this enables the fault tolerance functionality to manifest in a manner which will ensure that the operations will continue to take place uninterrupted and that the overall redundancy cost will be kept to a minimum.

The proposed IPFT model provides fault predictions by using data features which are associated with the resource usage in distributed edge environments. The IPFT monitors the resource consumption that takes place on each processing node in order to reveal, at run-time, insufficient processing capabilities that may result in potential QoS degradations. In case that the deployed resources cannot satisfy the increasing amount of demands within a specified time-frame, the IPFT will then trigger mitigation policies such as proactive node replication and task execution migration.  The multi-channel neural network part of the IPFT mechanism of a node takes into consideration the state of the other available nodes when predicting its future state. This way, in case of a predicted fault, it can perform task migration to the nodes that are already up and running, provided that there is enough computational capacity available. By doing so, we avoid immediately resorting to node replication which could lead to a cost and time deployment increase for setting up a new node. 

\begin{figure}
    \centering
    \includegraphics[scale=0.15]{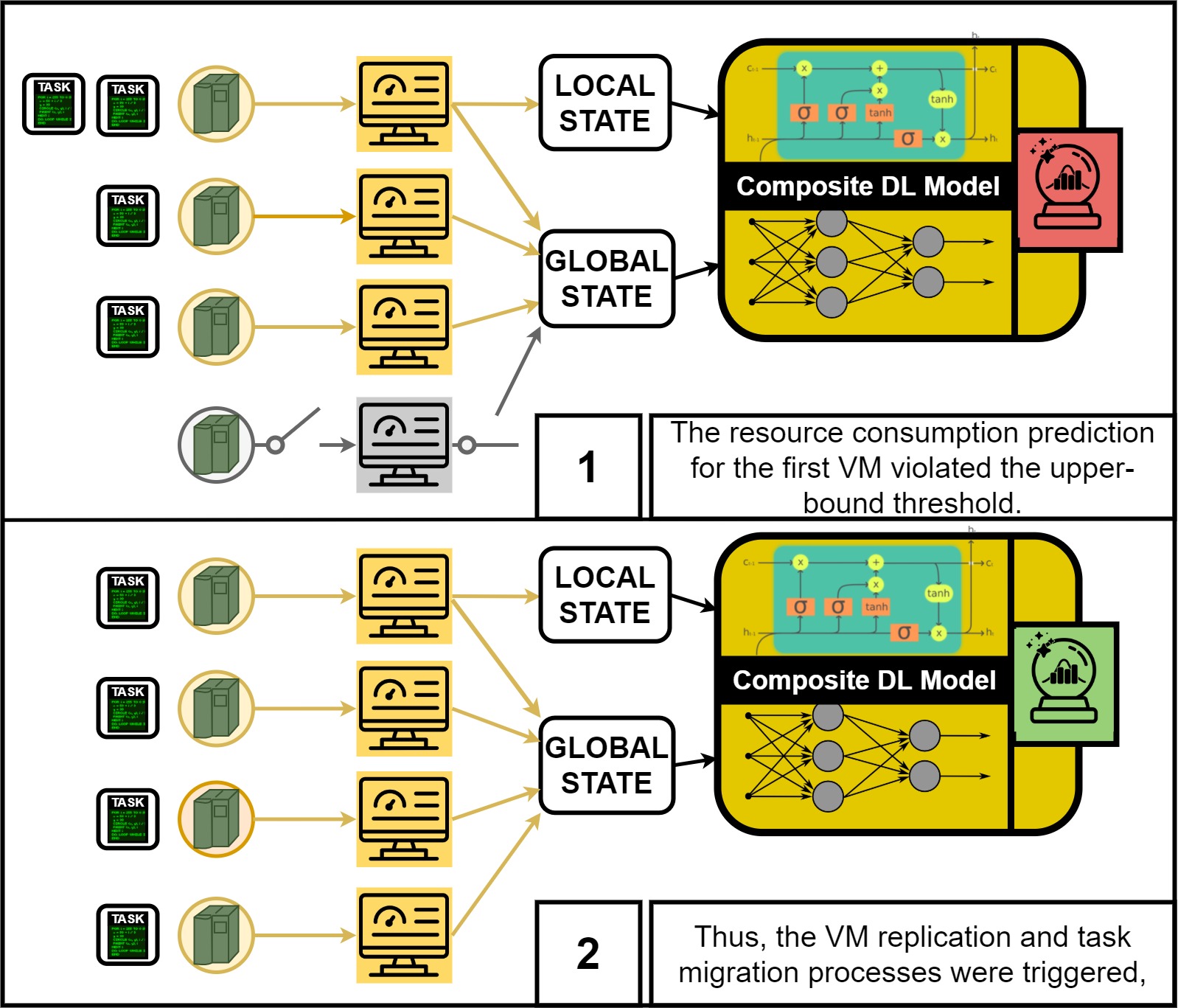}
    \caption{The IPFT Triggers Virtual Machine Replication \& Task Migration based on Resource Utilization Predictions.}
    \label{fig:FigX1}
\end{figure}

As mentioned before, the replication of a node requires a certain waiting period, which can have grave ramifications on the performance of the edge environment. The IPFT can predict the future needs for node replication in a time horizon longer than the replication time. Thus, the timely triggering of node replication processes and the corresponding task migration prevents the occurrence of the process faults. As illustrated in Fig. \ref{fig:FigX1}.1, when a specific processing node is predicted to present high resource utilisation, then a node replication process will be triggered (Fig. \ref{fig:FigX1}.2). This way, the future tasks to come will be accommodated by the new node, avoiding tasks rejections and long execution times.  

\begin{figure*}
    \centering
    \includegraphics[scale=0.15]{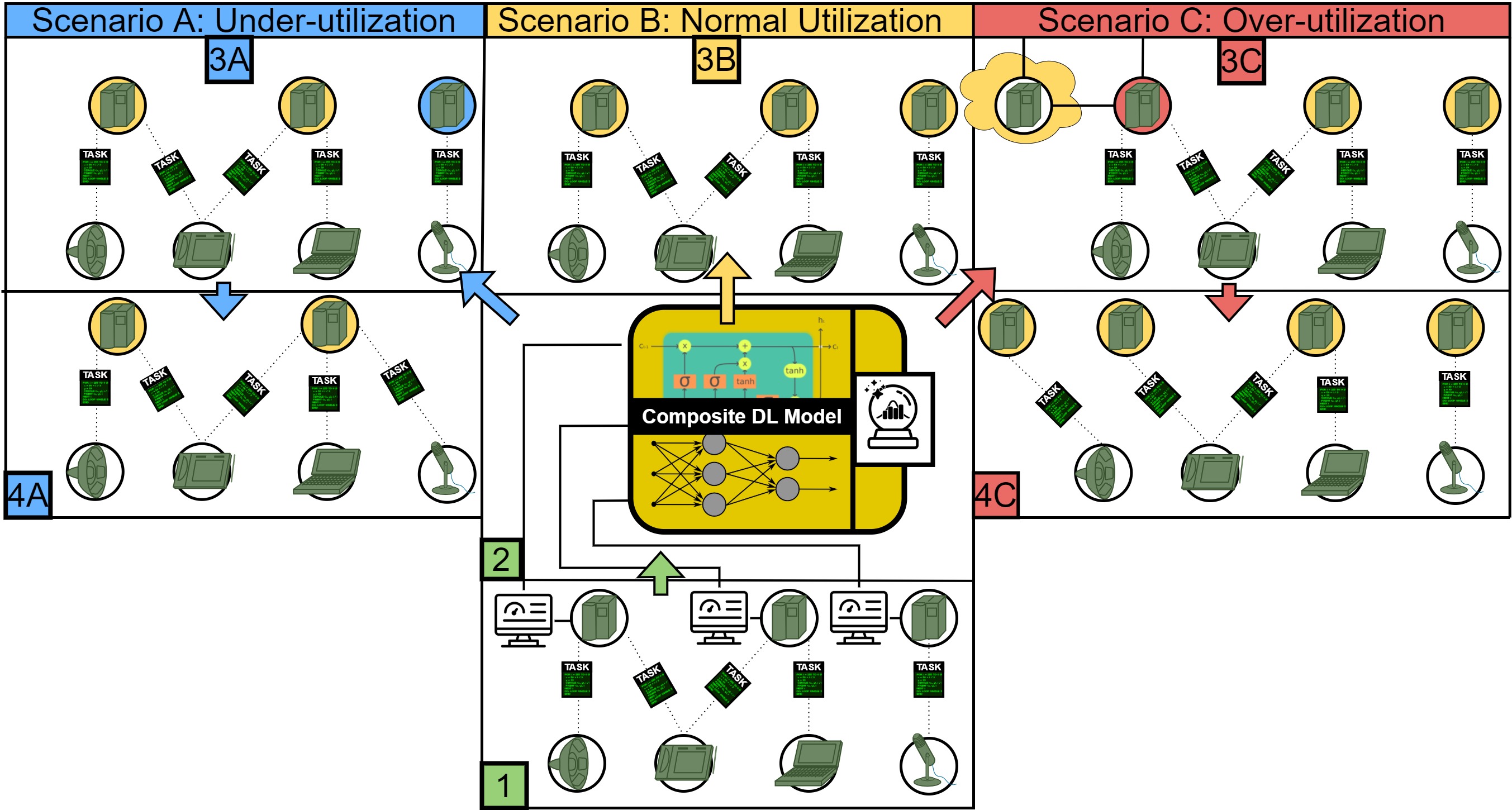}
    \caption{The four stages of the IPFT pipeline which span over three distinct scenarios.}
    \label{fig:FigX}
\end{figure*}

The IPFT operates in a manner which consists of 4 main stages, as one can see in Fig. \ref{fig:FigX}. In stage 1, it monitors the resource metrics of the processing nodes. In stage 2, it predicts the maximum resource usage that is expected to take place within a specified time-frame. With regards to the stage 3, there are three distinct scenarios which dictate how the rest of the operation shall be carried out. The stage 3 processes are carried out independently for each of the processing nodes. The three distinct scenarios and their interactions with the corresponding actions taken at stage 4 can be summarized as follows:
\begin{itemize}
  \item Stage 3A. If the resource usage prediction is lower than a specific lower-bound threshold, then the processing node is considered under-utilized and thus a decommission request for this specific node is issued. During the stage 4A, the tasks that were assigned to this node are redirected to alternative locations and the decommission process is completed.
  \item Stage 3B. If the resource usage prediction is between the lower-bound threshold and the upper-bound threshold, then the resource consumption rate is considered ideal and thus there is no need to perform any additional actions.
  \item Stage 3C. If the resource usage prediction is higher than a specific upper-bound threshold, then the processing node is considered over-utilized and thus  a replication request is issued. The stage 4C takes place after the creation of the new processing node. During the stage 4C, a fraction of the tasks that were assigned to the over-utilized processing node are redirected to the newly created one.
\end{itemize}

The IPFT works jointly with a workload balancing mechanism that gives higher priority to the nodes with low resource utilization predictions and lower priority to the nodes with higher resource utilization predictions. This is a common practice to many task offloading mechanisms that use different criteria to balance the workload \cite{8029252}. As an example the MinMin task offloading method prioritizes the smaller tasks to be executed in the nodes that will be available sooner. While the MaxMin task offloading method prioritizes the larger tasks to be executed in the processing nodes that will be available sooner. The size of the tasks is estimated based on the number of its million instructions or their estimated completion time. In this way, IPFT modifies the behaviour of the task offloading mechanisms taking into consideration the resource usage predictions for the migration of the tasks. 

\subsection {Threshold-based Decision Making}
Task migration and node replication occur when a resource usage prediction metric is higher than a specified threshold value. This type of threshold-based approach is used in many decision making mechanisms in  cloud/edge computing \cite{lin_threshold-based_2011}. The IPFT involves two thresholds. If the value of the resource utilization prediction is higher than the upper-bound threshold the IPFT invokes a node replication process. If the prediction value which corresponds to a specific processing node is lower than lower-bound threshold, then this particular processing node is turned off (e.g. for reducing the total energy consumption \cite{Avgeris_2022}). 

The appropriate selection of these two thresholds is integral to the performance of the IPFT. A high value in the upper-bound threshold will result in a system which is not sensitive and reactive enough to the workload fluctuations. While a low value in the upper-bound threshold will make the system to react and trigger unnecessary node replications. Similarly, the lower-bound threshold should be appropriately fine-tuned. A low value will make the infrastructure to continue using under-utilized processing nodes. A high lower-bound threshold will also turn off processing nodes that are necessary to the smooth operation of the edge infrastructure \cite{noauthor_automatically_nodate}.

In order to identify the optimal threshold values, we propose the use of a grid-search approach that iteratively tries sequential thresholds in order to converge close to the optimal values. These values maximize the fault tolerance evaluation metrics of reliability and maintainability, which will be extensively discussed in the upcoming experimental section. The selection of the threshold values is heavily dependant on the characteristics of each application and the underlying physical infrastructure. The literature provides recommendations in similar problems and mechanisms which, despite providing sub-optimal results, serve as valuable guidelines towards establishing some standards regarding how these bounds are chosen. \cite{thresholds}  

Given that the predictions are quite accurate and the threshold values are chosen optimally, the resulting system is expected to be highly robust and able to provide satisfying availability. For the best results, the resource utilization predictions should derive from the resource consumption behaviour of each individual processing node, while taking into consideration the overall resource consumption behaviour of the entire edge computing infrastructure. In the next section we will describe the deep learning model that predicts the resource utilization, as well as the way it simultaneously leverages the resource consumption metrics of each individual node and the entire edge infrastructure. 

\section{A composite Deep Learning Architecture for Resource Usage Prediction}
\label{Sec:RNN}
A composite deep learning network with the HBES model is proposed to provide accurate resource utilization predictions for the IPFT method. The composite deep learning network is designed to satisfy the particularities of the edge infrastructure and the resource usage metrics. Since resource metrics like CPU, RAM, disk, and bandwidth have sequential dependence, RNN can provide an appropriate type of neural layers. RNN combines the advantages of deep learning with the characteristics of time series forecasting. There are different types of RNN architectures and the two most prominent are the GRU and LSTM. Each individual processing node is examined separately for future possible process faults. However, in order to trigger the node replication, the deep learning model of each node should be aware of its own status and the whole edge infrastructure status. In this paper, the edge node which is examined is also called local and the whole edge infrastructure is called global. Because the local and the global status affect each other we propose the use of a composite deep learning model that combines the two in order to provide the local resource utilization predictions. The way that these two different sources of information are combined will be explained in the next subsection.

   \begin{figure}
    \centering
    \includegraphics[scale=0.13]{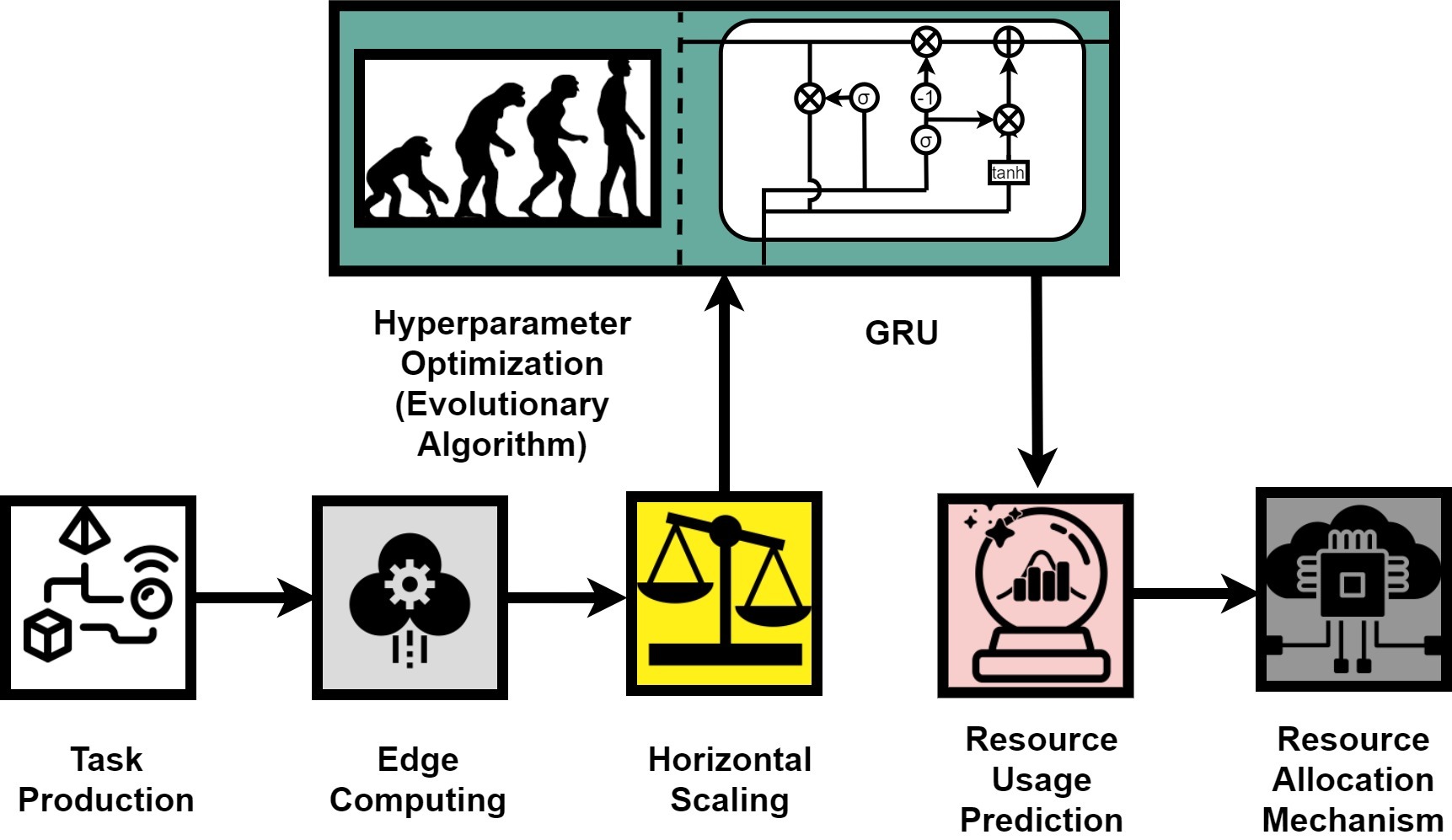}
    \caption{A Pipeline from Task Production to Advanced Fault Tolerance.}
    \label{fig:model_pipeline}
\end{figure}

The workflow of the resource usage prediction in an edge computing environment is depicted in Fig. \ref{fig:model_pipeline}. In the beginning, the edge devices generate tasks which are partially or fully executed to the edge computing nodes. During the task execution the nodes are monitored in order to keep the resource utilization metrics. The resource utilization metrics are provided to the composite deep learning model in order to predict the resource utilization in the next time horizon. Next, the resource usage predictions can be used by a thresholding method in order to trigger node replication and task migration. In the following subsections, we will describe the theory behind the key parts of composite neural networks, the RNN and the HBES for the resource usage prediction.

\subsection{Two Channel Architecture for Resource Usage Prediction} 
The most commonly used architecture, when it comes to neural networks, comprises of layers that are stacked one after the other in a serial manner. Input data is fed to the first layer, and one by one each of the layers transform the data and feed it to the next one. When building a neural network however, we have the option of putting layers or even series of layers in parallel, to take advantage of complex input structural properties. These parallel sections of the network are processing different parts of the data independently and subsequently concatenate their output to send it to the final section of the network. We decided to use an architecture, such as the one described above, utilizing two separate parallel series of layers (two channels) \cite{bodapati_joint_2021}. This allows us to better handle the local and global data metrics that are available.

The main advantage of using a multi-channel neural network architecture is the ability to use different kinds of layers in the input stage of the model. As an example, many use cases from the literature use a convolutional network and a feedforward network at the input stage to accommodate data having both numerical and image properties. Multiple channel neural network architectures have the advantage that they can group together the data that their properties present high correlation \cite{wang_two-channel_2019}. This happens by having different sequences of layers handling different kind of data.

A multi-channel neural network can be more time and resource-consuming to train and infer compared to a vanilla serial neural network, because of its more complicated structure. We also cannot properly evaluate the channels, or understand their individual effect when it comes to the final output of the model, since training is performed start-to-finish without giving us more details about the process.

RNN are a type of neural network that do particularly well when facing problems with sequential data. They are used to solve problems such as text prediction or voice recognition, and they are deemed effective when dealing with time series based problems as well. Consequently, we chose to include a RNN in our model as the first input channel, since it is going to handle the sequential data of a single node (i.e. the one that the model is actively trying to predict). The input is the monitoring measurements with an interval record of one minute. The channel consists of one or two RNN layers followed by a feedforward layer, and sends the results to the output part of the network.

For the second channel of the model, we chose to implement a feedforward network. This part is being fed with the global monitoring state, as well as a transformed timestamp feature that refers to the day of the week and the part of the day. It is obvious that the chance that a particular node to be overloaded in the immediate future is tied closely to the state of the other nodes that are collaborating with it to handle the requests. The channel of the feedforward network includes dense layers, that also have dropout layers in between. This architectural decision works as a regularization measure that stops our model from overfitting during training.

Once the channels handle the input part of the model, their concatenated results are given to the output section of the network. The goal of this combination is to take into account the processed data and output a prediction for the resource utilization of a particular node, in a sufficient time horizon after the state/last input we fed to the model. This is a feedforward neural network as well, with dense and dropout layers until the output layer. The choices regarding the merging of the two channels and the concatenation size of the input section, are handled by the hyperparameter selection algorithm. 

In order to train itself on a dataset, the model tries to predict the future state of a particular node based on the complex input referring to both the local node in question and the global overall state. Once the model outputs a prediction, it compares the prediction to the actual output value utilizing a loss function. Loss functions such as MAE or RMSE, are a way to tell how good or bad a regressive model is at predicting the correct values. Based on the error calculated, the model shifts its weights and parameters using a variant of gradient descent, back-propagating through the network affecting all layers. A function called optimizer is describing how the procedure of weight-shifting/network-optimizing takes place. Both the architecture, and the training of a neural network provide us with many options and hyperparameters to tweak. In order to ensure a quick and optimal choice of hyperparameters for our model we implemented the HBES algorithm that will efficiently make these decisions.

\subsection{ Recurrent Neural Networks for Time Series Data}
Artificial neural networks can be defined as function approximators, mapping lower level data representations to higher and disentangled data representations. RNN \cite{shiva_prakash_survey_2019} is a type of artificial neural networks, which facilitate dynamic temporal behavior, capture data sequences, and maintain the previous input states.The RNN architectural paradigm is based on various neuron-like nodes organized into successive layers, where each node is connected with nodes of the next successive layer and also has recurrent connections. Utilizing this particular concept, information regarding previous data inputs is allowed to affect future outputs, thus making RNN a solid option for time series modeling while taking into account contextual information. By monitoring edge computing infrastructures, we gather sequential data and predict the future resource usage metrics with RNN, based on their current and previous values. 

The main problem that RNN encounters is the vanishing gradient problem. This problem emerges during the training stage of the RNN, when the gradients become vanishing small preventing the weight updates of the RNN. Various gate-related architectures have been introduced in order to tackle the vanishing gradient problem. Through the use of gates, the network is able to properly maintain relevant information and to successfully pass it down to the next time-steps. The two most notable ones are the LSTM networks and the GRU \cite{shen_deep_2018} networks.

\subsection{Long Short-Term Memory}

The LSTM network architecture was created to tackle the problem of vanishing gradients in RNN. The importance of using this complicated architecture can be highlighted by pointing out how much context can offer in finding a solution. When dealing with a time series data problem, such as predicting resource usage, we can get much more useful information if we look at historical data of our machines' usage, rather than just glancing at their current state. This way, we can better understand concepts like trend, which can only be explained over time. LSTMs, just like regular RNNs, utilize the hidden state to connect the consequent nodes so as to enable better understanding of temporal data. However, they also use a cell state, which is another connection between the nodes. Each LSTM cell can read from the cell state, write to it or reset it via the use of gates.

There are three gates in total, each activated by a sigmoid function. This ensures that the model remains differentiable, since sigmoid offers smooth curves in the range of 0 to 1. Each one of the gates take as inputs the actual input as well as the hidden state of the previous timestep.

In addition to the gates, a vector called $\overline{\text{C}}$ is responsible for carrying the candidate information that can be added to the cell state. $\overline{\text{C}}$ utilizes a tanh layer, which is in charge of limiting the vanishing gradient phenomenon. To this extent, the cell information can be kept longer without vanishing it. The way this is achieved is by keeping the gradients zero-centered, between the values of -1 and 1.

The input gate is handling incoming data, and controlling whether the memory cell should be updated. It is applied to $\overline{\text{C}}$, and the result is then added to the cell state. The sigmoid activation of the gate is used to either mitigate or enhance the effect that the new information should have on the cell state.

The forget gate is the entity that is responsible for selecting the information that is deemed less important, and it removes it from the cell state by soft-resetting its values. Additionally, by utilizing the sigmoid function, it produces a scaled output for every value that is saved in the cell state.

Finally, the output gate is the final layer before the new hidden state is produced. It uses the sigmoid function as a filter to be applied to the cell state after it goes through a tanh layer first. After this process is completed, both the hidden state and the cell state compose the output of the LSTM cell which will be inputted to the next time-step.

\subsection{Gated Recurrent Units}
GRU and LSTM are similar as both of them manage to prevent the vanishing gradient problem by utilizing gate structures. What sets them apart is the fact that GRU combines the forget gate and input gate to form a single update gate. By reducing the number of gates involved, GRU is able to provide less complex structures and thus, be more computationally efficient when compared to LSTM. At the same time, GRU manages to perform equally well.

GRU networks also facilitate the hidden state mechanism which connects one unit of the network to the next, thus allowing the manifestation of dynamic temporal behavior in a similar manner. Each GRU unit is indicative of a specific time-step that facilitates the transfer of important information through the time continuum. Furthermore, it contains two distinct gate structures. The first one is referred to as the reset gate while the second one is referred to as the update gate. They both bear sigmoid layers which provide smooth curves in the 0 to 1 zone, thus ensuring that the model will remain differentiable. By squishing the values between 0 and 1, the sigmoid activation also helps the network to learn which data is important or not and then accordingly keep it or forget it. 

\begin{figure}
    \centering
    \includegraphics[scale=0.07]{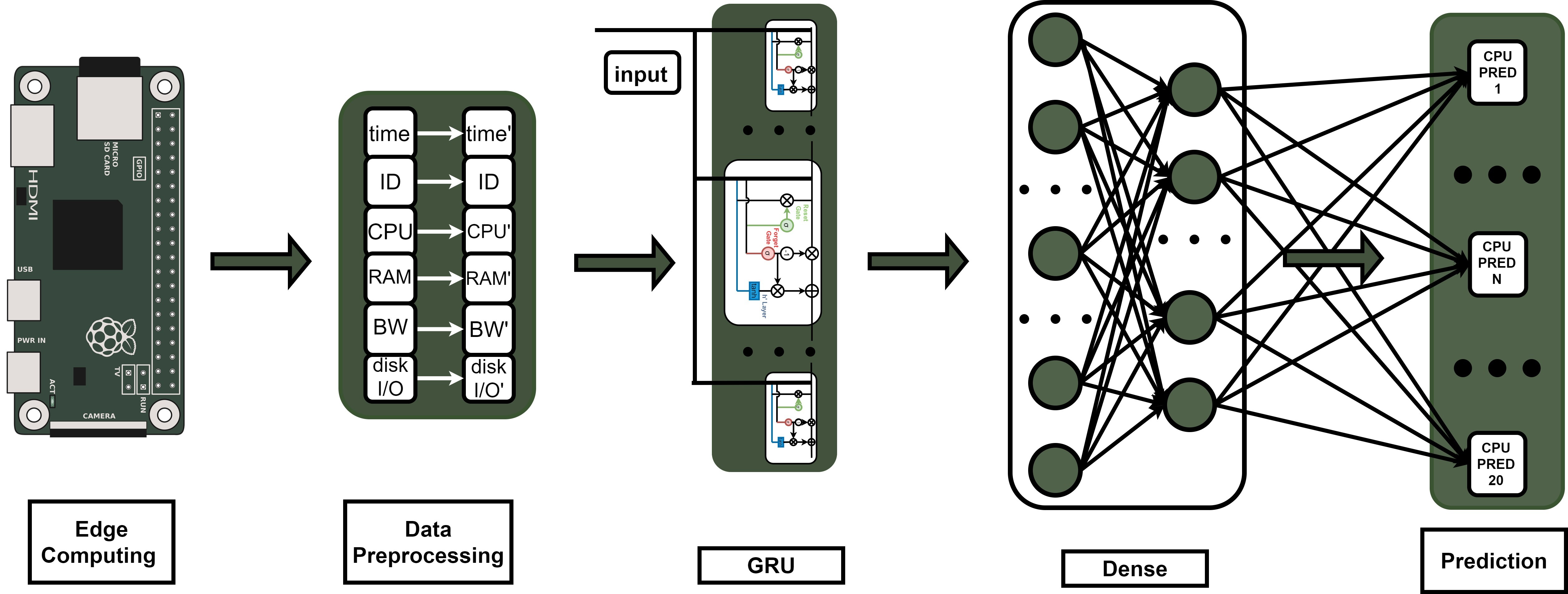}
    \caption{Integrated RNN in the resource usage prediction.}
    \label{fig:GRU}
\end{figure}

In order to contextualize the GRU paradigm in accordance to edge computing, we input vectorized representations with information such as the perspective timestamps, the resource utilization of CPU, RAM, bandwidth, and disk through the data preprocessing as illustrated in Fig. \ref{fig:GRU}. The functionality of GRU networks is carried out in the form of the following steps. As explained before each GRU uses a reset gate and an update gate. Each of these gates has two weight matrices. The first one corresponds to the input while the second one corresponds to the hidden state. The reset gate of GRU is responsible for deciding how much of the past information shall be forgotten. Much like in the case of LSTMs, the first step is to multiply the input and the hidden state by their corresponding weights. The sum of the multiplication results is then passed through a sigmoid layer.

The update gate is in charge of determining how much of the information gathered over the previous time-steps needs to be passed along for future use. In this regard, its behavior is quite similar to the one of the reset gate. The first step requires the multiplication of the input and the hidden state by their perspective weights. The hidden state entails information derived from the previous $t-1$ units. Then, the multiplication results are added together and passed through a sigmoid layer. The output of the update gate will be referred to as $u$.

The next step is to create a candidate new hidden state. Similarly to the reset and update gates, there are also two weight matrices involved. The first one corresponds to the input and the second one corresponds to the hidden state. The first step towards creating a candidate new hidden state is to multiply the input by its corresponding weights. The second step is to calculate the Hadamard product in an element-wise manner between the hidden state and the output of the reset gate. This process is essential for deciding how much of the information gathered during the previous time-steps will be removed. The Hadamard product is then multiplied by the weights of the hidden state. The results of the two multiplications are then added together. The sum is passed through a tanh layer, which minimizes the effects of the vanishing gradient phenomenon. This is performed by distributing the gradients in a sufficient manner, within a zero-centered range. Thus, it enables the information to flow longer without vanishing. The product of the operations so far is the candidate new hidden state will be referred to as $h'$.

In order to get the updated hidden state, the first step required is to perform element-wise multiplication to the output of the update gate and the hidden state. The second one is to perform element-wise multiplication to the $h'$ and the product of $1-u$. The updated hidden state is the sum of the two multiplication products. The updated hidden state is then carried over to the next GRU unit, which corresponds to the next time-step.

\subsection{Evolutionary Strategy}
Evolutionary strategy  \cite{hansen_evolution_2015} belongs to the category of evolutionary algorithms that are population-based metaheuristic optimization approaches inspired by the principles of biological evolution. The formulation of evolution strategy is based on successive iterations of mutation and selection over a population of candidate solutions. The candidate solutions, also named individuals, are initialised in random positions in an n-dimensional space and move toward positions that minimize an objective function. These dimensions are the numerical GRU-RNN hyperparamaters that should be optimized.

For the needs of hypertuning GRU-RNN, we used its numerical hyperparameters as the search space of the evolution strategy and the mean squared error of the candidate RNN as the fitness function. In each iteration, a number of RNN are trained and evaluated with the mean squared error, whereas the most accurate of them are mutated to the next iteration. The mutation is a stochastic process based on a normal distribution that introduces variations in the best fit individuals of each iteration. In the beginning, the exploration for different candidate solutions is intensive, making stronger mutations towards new areas of the search space. In each iteration the exploration decreases and the exploitation of the best fit individuals increase using a self adaptation control variable. This means that the mutation introduces strong variations in the first iterations and the variations decay as the evolution progresses in order to converge to a close to optimal RNN architecture.

Hypertuning deep learning models with evolution strategy in contrast with other evolution algorithms, like genetic algorithms, has the advantage of not re-combining different neural network topologies that may have significant discrepancies in their phenotypes. This happens because the crossover of the genetic algorithm has the difficulty that the parents may have different architectures that cannot be unified in their offspring.  A typical example is if the one parent is a 2 layered LSTM-RNN followed with 6 dense layers and the second parent is a 2 layered GRU-RNN followed with 4 dense layers. Thus, the phenotypes of LSTM and GRU cannot be smoothly recombined. On the other hand, evolution strategy is based only in  the selection and the mutation which smoothly lead the evolution process. Specifically, the mutation operations introduce variations into the survived candidates providing the opportunity to test neighbour solutions that may lead to an improved fitness value.

\subsection{ Bayesian Optimization}
Bayesian optimization \cite{frazier_tutorial_2018} is widely used to estimate hyperparameters in machine learning and deep learning models. It was an obvious option for the searching process in the categorical dimensional space, in order to find the close to optimal nominal hyperparamaters of the RNN. Bayesian optimization iteratively requests new observations of the search space with an acquisition function and estimates the objective function with a surrogate function. The increase of the Bayesian optimization observations gives a higher probability for the global optimum location. Nonetheless, we should take into consideration that the number of observations are finite and computationally expensive so the smart search process should select points that maximize the probability to find a new optimal following an exploration vs. exploitation trade off. 

The surrogate function approximates the objective one and is updated every time the objective function is evaluated in the new candidate points. The acquisition function decides where to sample next in the iterative process of Bayesian optimization, finding the points that maximize the expected improvement. The expected improvement is a function of two components. The first estimates the regions that the surrogate function has optimal points and the second estimates the regions with high prediction uncertainty that have not explored efficiently yet.

\subsection{Hybrid Evolution Strategy with Bayesian Optimization}
The hyperparameter optimisation for a composite neural network is a prominent challenge as it includes the important architectural decisions for a close to optimal topology. The HBES constitutes an innovative, holistic and unified approach for hypertuning by merging the evolution strategy and Bayesian optimization methodologies.  The evolution strategy is responsible to evolve a population of candidate deep learning models based on their numerical hyperparameters and each individual candidate solution estimates its nominal hyperparameters with the Bayesian optimization as it is described in Algorithm \ref{Alg:Algo1}.

\begin{algorithm}
\caption{Hybrid Bayesian \& Evolution Strategy }
\label{Alg:Algo1}
\begin{algorithmic}

\STATE \textbf{Step 1:} Initialization of Evolution Strategy 

\hspace*{3mm} Set the starting search point of the algorithm. \\
\hspace*{3mm} Usually $a_1$=[0.5, 0.5, ..., 0.5] since we have \\ 
\hspace*{3mm} already scaled our hyperparameter options \\
\hspace*{3mm} down to [0,1]

\hspace*{0mm}\textbf{Step 2:} for i = 1, 2, ..., $n_{pop}$: \\ 
\hspace*{3mm} i) add some random noise to the search point \\
\hspace*{3mm} ii) Scale back from [0,1] to the hyperparameter \\
\hspace*{3mm} search space to create the ordinal \\
\hspace*{3mm} hyperparameter values for the network to be \\
\hspace*{3mm} trained \\
\hspace*{3mm} iii) Bayesian Optimization with GP\\
\hspace*{5mm} 1) Apply a Gaussian Process prior on $f$\\
\hspace*{5mm} 2) Observe $f$ at $n_0$ points according to an \\
\hspace*{8mm} initial experimental design\\
\hspace*{5mm} 3) Initialize $n=n_0$\\
\hspace*{5mm} 4) Repeat while $n \leq N$ \\
\hspace*{7mm} a) Update the posterior probability \\
\hspace*{7mm} distribution on $f$ using all available \\
\hspace*{7mm} data\\
\hspace*{7mm} b) Let $x_n$ be a maximizer of the \\
\hspace*{7mm} acquisition function over x.\\
\hspace*{7mm} c) Observe $y_n=f(x_n,x_i(t+1),v_i(t+1))$\\
\hspace*{7mm} d) $n \leftarrow (n+1)$\\
\STATE \textbf{Step 3:} Sort the results and its corresponding\\
\hspace*{11mm} hyperparameters \\ 
\STATE \textbf{Step 4:} Calculate the new search point by\\
\hspace*{11mm} averaging the points of the $top_n$ networks \\ 
\STATE \textbf{Step 5:} Go to Step 2 until desired number of\\
\hspace*{11mm} iterations is completed\\

\end{algorithmic}
\end{algorithm}

The numerical hyperparameters are the number of recurrent layers and feedforward layers, the number of neurons for each layer, the lookback, epochs, the batch size, and the percentage of dropout and learning rate. The nominal hyperparameters are the type of the neural layers, the activation functions and the optimizers. The gained knowledge of the nominal hyperparameters is universal through the population and updated by all the individuals over the generations. The ultimate goal of HBES is through the Bayesian evolution process to converge to a close to optimal solution and to train deep learning models that can predict timely and accurately the resource utilization of the next timesteps.

\begin{table*}
\centering
\caption{Comparison of single-output \& multi-output prediction methods of resource usage metrics.}\label{tab_eval}
\begin{tabular}{|l|l|l|l|l|l|l|l|l|}
\hline
Method  & RMSE & MAE  & \multicolumn{2}{c|}{
 \begin{tabular}{c}
CPU-1 (\%)  \\ \hline \begin{tabular}{l|l} RMSE & \space MAE  \end{tabular}
 \end{tabular}

} &  \multicolumn{2}{c|}{
 \begin{tabular}{c}
RAM-1 (\%)  \\ \hline \begin{tabular}{l|l} RMSE & \space MAE  \end{tabular}
 \end{tabular}
} 

& \multicolumn{2}{c|}{
 \begin{tabular}{c}
Infer. Time  \\ \hline \begin{tabular}{l|l} Single & \space Batch  \end{tabular}
 \end{tabular}
}

\\
\hline

HBES-GRU  &  \underline{0.0641} & 0.0276 & \underline{15.918} \space \space  \space\space & \underline{12.815} & 1.694\space\space\space\space\space \space \space  & 0.580 &   \space \space 0.033 \space\space\space\space \space & 0.038\\
GA-LSTM  & 0.0674 & 0.0338 & 16.099 & 12.838 & 1.746 & 0.917 & 0.020 & 0.024\\
Keras-Tuner & 0.0785 & 0.0377 & 16.291 & 13.290 & 2.631 & 0.818 & 0.042 & 0.042\\
AUCROP  & 0.0814 & 0.0414 & 17.235 & 14.009 &2.480 \space \space \space & 1.482 & \space \space \underline{0.004} \space & 0.011\\
XGBoost & 0.1139 & 0.0599 & 16.457 & 13.569 & \underline{1.515} & \underline{0.472} & 0.060 & \underline{0.010}\\
Auto-sklearn & 0.1055 & \underline{0.0243} & 52.659 & 17.856 & 1.546 & 0.526 & 0.263 & 0.572\\

\hline
\end{tabular}
\end{table*}

\section{Experimental Evaluation}

To evaluate the IPFT methodology, we make two types of experiments. First, we experimentally evaluate and compare the applicability of RNN with the HBES against state of the art methods using a real dataset. This dataset is constructed by monitoring Raspberry Pi's in an edge computing infrastructure. Next, we leverage the resource utilization prediction model in order to develop an IPFT mechanism that takes intelligent replication and migration decisions in a sufficient time before the process faults occur. The sufficient time in the context of intelligent replication and migration involves the service deployment time of the processing nodes and the scheduling of the new tasks on that node. The experimental evaluation of IPFT took place in a seven days edge computing simulation using the CloudSim Plus framework \cite{silva_filho_cloudsim_2017}

\label{Sec:Experimental}

\subsection{Experimental Evaluation in Resource Usage Prediction}
The edge infrastructure we used includes Raspberry Pi3 as processing nodes with a 64-bit quad-core ARM Cortex-A53 at 1.4GHz, loaded with Raspbian operating system which is a version of Debian Linux. The dataset constructed by a monitoring tool implemented in Python 3 using the libraries psutil \cite{psutil} and GPUtil \cite{gputil}. We monitored the real time usage of CPU, RAM, disk and bandwidth in one second time interval. 

The deployed application was a natural language processing text classification. The use case was to make the text classification on an edge computing environment, locally, close to the text owners and not in cloud computing infrastructures for privacy issues. The reason for this choice is that the text owners did not agree for their texts to be transferred and processed in remote servers. In order to control the application remotely and take the resource usage datasets we used the SSH protocol, but we did not have the privileges to access the processed texts.

\subsubsection{Model Implementation and Frameworks for Comparison}
The HBES model and the RNN multi-output regression model are implemented in Python 3 using the frameworks NumPy, pandas, statistics, Scikit-learn, SciPy, Scikit-Optimize, TensorFlow 2 and Keras. The environment we used for the training and the evaluation of the model was the Jupyter notebook of the Google Colaboratory. The experiments’ source code is available for any kind of reproduction and re-examination in our GitHub repository \cite{stsanakas}. In this experimental setup we used the HBES with GRU as RNN and compared the results with a time series baseline approach, the machine learning meta-model for resource usage prediction AUCROP, the Auto-sklearn, the XGBoost, our previous LSTM with genetic algorithm model (GA-LSTM) \cite{violos_lstm} and the Keras-Tuner.

\subsubsection{Evaluation Results and Discussion}
Our initial time-series analysis provided results which indicate positive correlations for lags in a value range from 1 to 22. This finding confirms the strong self similarity property that the sequences of resource usage metrics have. Afterwards, using the ARIMA forecasting model we evaluated the resource metrics predictions. For instance, the CPU RMSE was 18.474. After comparing  the results of the statistical models with the ones of the machine learning and deep learning approaches we found that the latter had an improvement that surpasses the 20\% RMSE in most cases. Because of that, we decided to focus our research on the machine learning and deep learning models.

  \begin{figure}
    \centering
    \includegraphics[scale=0.7]{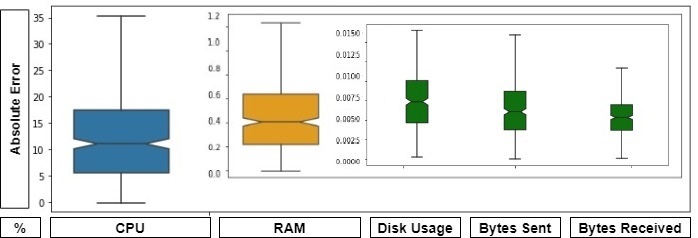}
    \caption{GRU-RNN with HBES prediction errors of resource usage metrics.}
    \label{fig:boxplot}
\end{figure}
 
Table \ref{tab_eval} summarizes the experimental results. The first two columns provide the aggregated RMSE and MAE including all testing values of the devices, and the resource metrics. For RMSE, which gives an extra penalty to predictions with significant errors, we can see that HBES-GRU had the best performance. In the column entitled MAE, we see that the two best models are the auto-sklearn and HBES-GRU. Their prediction errors are very close and they have a significantly better performance compared to other models. 

CPU-1 and RAM-1 columns represent the RMSE and MAE for the processing edge node which had the least accurate predictions in the infrastructure. In addition, Fig. \ref{fig:boxplot} illustrates the 25th and 75th percentile, the median, the min and the max of the error value metrics. These metrics include CPU, RAM, disk usage, and bandwidth in terms of the bytes sent and received. Regarding the disk usage and the bandwidth the prediction errors were insignificant. This is not only due to the ability of the HBES-GRU to provide accurate predictions, but due to the small fluctuation in these two resource metrics as well. The fluctuation in CPU is much greater than in RAM and HBES-GRU captures in a better way the various changes when compared to the other models. XGBoost has better performance than HBES-GRU in RAM. This may be justified by the ensemble structure that XGBoost has. XGBoost can build specific decision trees for the residuals of RAM and target on its slow change behaviour. 

Lastly, we see the inference times of the models which are required in order to provide a single prediction or a batch with one hundred predictions. All time measurements are in seconds. All the inference times, except from Auto-sklearn, are within a range from 11 to 60 msec. These inference times indicate that resource usage prediction is a rather fast process which can be incorporated in time sensitive applications. In this research we did not compare the training times because we wanted to make an exhaustive smart search in the hypothesis space and see the limits of accuracy that the different models can achieve. It is worth noting that we have made experiments using a wide range of time-frames. However, the measurements in Table \ref{tab_eval} are produced using a 10 minute time-frame. We chose to illustrate this time-frame because it is close to the actual time which is required for the deployment time of a node. From the results we see that even if GRUs are simpler in their structure compared to LSTMs, due to their lack of a dedicated output gate, they had slightly better performance. Furthermore, the conducted experiments enabled us to reach to the following conclusions:
\begin{itemize}
  \item In most metrics the deep learning models (HBES-GRU, GA-LSTM, Keras-Tuner) have better performance than machine learning models (AUCROP, XGBoost, Auto-sklearn).
  \item The evolutionary algorithms for hypertuning (HBES-GRU and GA-LSTM) have better performance compared with the simple Bayesian optimization (Keras-Tuner).
  \item One can witness a significant improvement when using the hybrid Bayesian and evolution strategy approach instead of simple genetic algorithms.
\end{itemize}

\subsubsection{Convergence of Hybrid Bayesian Evolution Strategy}
The convergence and the location of the global optimum are two of the most important topics in the domain of evolutionary algorithms. The convergence means that as the population evolves, the individuals go closer to the optimal solution shrinking their divergence. However, we cannot be sure if the convergence points in the genotype space constitute a global or local minimum. For this reason, the HBES algorithm in the beginning of the evolution process expresses a strong variance in the mutation which decays over iterations. Concurrently, we keep the best genotype found over all the iterations.

 \begin{figure}
    \centering
    \includegraphics[scale=0.40]{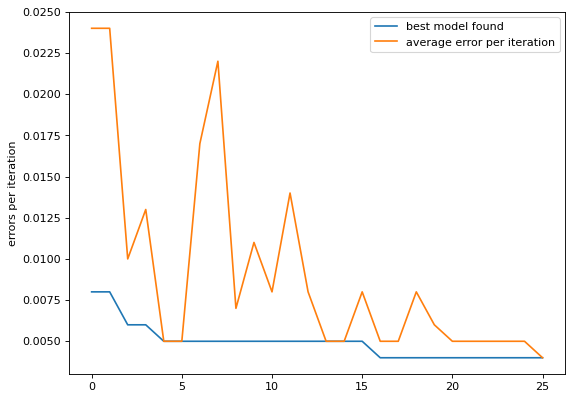}
    \caption{The convergence of HBES for close to optimal RNN.}
    \label{fig:convergence}
\end{figure}

The convergence of HBES is illustrated in Fig. \ref{fig:convergence}. We observe that in the beginning the average population error per iteration fluctuated strongly. In some iterations it is trapped in local minima as example between the iterations six to eleven. In some other iterations it lays in plateau regions, such as between iterations twenty to twenty five. Yet, using the mutation the individuals eventually escape from the plateau regions and the local minima and move towards close to optimal regions. 

These close to optimal regions in the genotype space are decoded to the close to optimal GRU-RNN architectures in the phenotype space. These GRU-RNN architectures provide the most accurate resource usage predictions for CPU, RAM, disk, and bandwidth usage in an edge computing infrastructure.

\subsection{Experimental Evaluation in Proactive Fault Tolerance}
The promising experimental results of the HBES with RNN in regards to resource usage prediction, motivated us to continue the experiments in order to evaluate its applicability as a proactive fault tolerance mechanism. Specifically, we used the HBES to make a smart search in candidate two-channel deep learning models and applied the thresholding method as described in Section IV.

\subsubsection{Experimental Simulation}

The composite deep learning model was integrated in an edge simulation of CloudSim Plus. We simulated an edge infrastructure that consists of a set of nodes, 5 available to us by default, and another 15 that can be activated for intelligent replication when needed. We simulated the local and global resource monitoring process, measuring CPU, RAM and bandwidth values, and saving those values every 60 seconds (time-step). The task offloader of the infrastructure was receiving incoming traffic and was assigning each task on a node, based on the following scheduling algorithms: RoundRobin, MinMin, and MaxMin.

The local and global resource usage metrics are being monitored and then fed to the IPFT mechanisms of each processing node. During every single time-step we use the monitoring data in order to formulate the appropriate data representations, featuring the past time-series measurements of a single node, as well as the state of the infrastructure as a whole. The input is then fed to the composite deep learning model, enabling it to make predictions of resource usage for every node in a time horizon of 10 minutes.  In this experimental set up we made the assumption that the preparation time for the infrastructure to assure its availability and robustness to faults is 10 minutes.

The simulation lasted for seven days and the tasks were generated by a mixture of Gaussian probability distributions that simulate a realistic application workload behaviour \cite{minh_realistic_2010}. The processing nodes simulated the processing capabilities of Raspberry Pi's. We defined a process fault if the time execution of a task lasts more than one second. The selection of one second is a reasonably acceptable latency for several data analytic applications \cite{big_data_fault}. Trying different latency times for the process faults, we noticed that the IPFT performance was better than the reactive approach. In the reactive fault tolerance approach, a node replication is triggered in case of a fault is taking place. In Table II, as we will thoroughly discuss in the next section, we compare the IPFT mechanism to the reactive fault tolerance approach.

\subsubsection{Fault Tolerance Evaluation Metrics}
In order to evaluate the performance of the IPFT mechanisms, we used a set of fault tolerance evaluation metrics \cite{ataallah_fault_2015}. Mean Time To Failure (MTTF) is defined as the expected time for a failure to occur given that the system functions properly. MTTF is an evaluation metric which corresponds to the overall inability of the edge infrastructure to operate properly and thus, it is calculated by taking into consideration the number of faults regardless of the actual processing node that failed. Mean Time To Repair (MTTR) is defined as the expected time required to repair the system after a failure occurs. For the MTTF the higher values are the better and for MTTR the lower values are the better. These evaluation metrics are calculated in terms of seconds. 

Two additional Fault Tolerance evaluation metrics are the reliability and maintainability. Reliability refers to the ability of an edge infrastructure to run continuously without any failure. Maintainability refers to how easily a failed system can be repaired. Both reliability and maintainability are numbers with no units and higher values mean better performance.

\subsubsection{Evaluation Results and Discussion}
The experimental results are summarized in Table II. We compared the IPFT mechanism to the Reactive Fault Tolerance (RFT) approach. The RFT approach performs node replications after a fault occurs. Regarding the task offloading algorithm we used the Round Robin (RR) \cite{alhaidari_enhanced_2021}, the MinMin and MaxMin \cite{derakhshan_optimization_2018}. The experimental result show the superiority of IPFT compared to the RFT in all evaluation metrics. In addition we see that the outcomes are significantly affected from the task offloading mechanism. This happens because the task offloading algorithms also integrate a workload balancing methodology with different criteria as we will discuss in the following paragraphs.

The number of generated tasks in all experimental setups was close to 1,500,000 with some parts of the day to have an intensive task generation (i.e., 11:00 -13:00), while some other parts of the day a small number of tasks (i.e., 02:00 - 04:00). We simulated this task generation behaviour because it is close to the activity of many user applications during a day. In this way, we observed that the infrastructure made intelligent replications during the parts of the day with an increased task generation. Respectively, the infrastructure turned off the edge nodes when the IPFT mechanism predicted an under-utilization of the processing nodes. Sometimes there were some sudden spikes or drops in task generation and resource utilization, but the infrastructure using the IPFT mechanism could dynamically and timely adapt. 

In Table II, we can see that in RR, MaxMin and MinMin the MTTF in IPFT has been increased compared to the RFT. This means that leveraging the resource usage predictions, faults occur more sparsely and rarely.  We can see from the MTTR metric that in the event of a fault, the infrastructure will recover very quickly, scheduling the new tasks in processing nodes with low resource utilization. The reliability metric shows that by using the IPFT, the edge infrastructure can provide the expected results up to 93\% of the simulation length, even during the stressing time periods of the simulated days. The significant improvement noticed for the maintainability metric, declares that even if a fault occurs, the IPFT will increase the robustness of the edge infrastructure. In other words, the IPFT will take timely the right measures by triggering node replication and task migration, in order to reduce the likelihood of subsequent fault occurrence.

A fault is recorded taking into consideration all the edge nodes that are currently active. This means that the MTTF value of 13.309 seconds in IPFT MaxMin includes the faults of different edge nodes. In addition to that, some generated tasks had a large number of million instructions that would have provoked a fault because of the CPU unavailability in the processing nodes. In this case, we wanted to know how these tasks affect the MTTF and MTTR. From the analysis of the results we saw that the variance of the task size is the reason that we see that the three different task offloading mechanism have different performance. In particular, the MaxMin algorithm gives higher priority in big tasks, thus we see a significantly better MTTF metric.

During the simulation we examined the IPFT decisions and how the edge environment operates. The simulation confirmed that the infrastructure takes advantage of the timely decision to trigger proactive actions, such as intelligent node replication and task migration before the amount of tasks overwhelms the processing nodes. This can be particular important for the infrastructure provider as it can save cost and energy, by shutting down nodes when they are no longer needed. Additionally, the provider can achieve a smoother flow of on-time completed tasks, avoiding crashes and minimizing QoS deterioration. 

\textcolor{black}{In regards to the actual cost of implementing the proposed IPFT paradigm, the consumption of computational resources was $3.2\%$ greater when compared to the reactive approach. Furthermore, the burden imposed on the network infrastructure was about $10$ bytes per second, due to the information flow that derived from the need of the prediction model to have access to the ongoing resource usage. Finally, the incorporation of the prediction mechanism used around 125MB of RAM and increased CPU consumption by around 47\% for an average of 2250ms on an Intel Xeon E312xx CPU. When contemplating the substantial benefits provided by the IPFT approach, we believe that the overall implementation cost is justified and quite reasonable. }

\begin{table}[]
\centering
\caption{Evaluation of Reactive \& Intelligent Proactive Fault Tolerance Methods}\label{tab_eval_FT}
\begin{tabular}{|l|c|c|c|c|}
\hline
                   & MTTF   & MTTR   & Reliability & Maintainability \\ \hline
RFT RR      & 2.864  & 19.657 & 0.741       & 0.048           \\ \hline
IPFT RR             & 9.506  & 3.343  & 0.904       & 0.230           \\ \hline
RFT MinMin  & 8.733  & 36.169 & 0.897       & 0.026           \\ \hline
IPFT MinMin         & 8.919  & 5.656  & 0.899       & 0.150           \\ \hline
RFT MaxMin  & 3.721  & 24.239 & 0.788       & 0.039           \\ \hline
IPFT MaxMin         & 13.309 & 7.425  & 0.930       & 0.118           \\ \hline
\end{tabular}
\end{table}

\section{Conclusion}
\label{Sec:Conclusion}
In this paper, we proposed a proactive Fault Tolerance mechanism in an edge computing infrastructure based on the resource usage predictions. In the beginning, we discussed and experimentally evaluated the use of 
RNN for the resource usage modelling. We developed a composite deep learning model that leverages in two channels the resource usage metrics of the local processing nodes and the infrastructure as a whole. We also designed a hypertuning algorithm that combines evolution strategy with Bayesian optimization and surpasses commonly used hypertuners like Keras-Tuner and other state of the art machine learning models. Last but not least, we presented how a proactive fault tolerance mechanism can leverage the resource usage predictions triggering node replication and task migration. 

Our experiments and results corroborated the efficiency of our proactive fault tolerance methodology, the applicability of RNN and the two-channel architecture for the resource usage prediction. The limitation of our work is that we have not yet holistically worked with other types of faults like network faults, physical faults and service expiry faults. Our future work is to make data analysis and find the execution patterns in the edge resources that are related with these types of faults. In addition, we want to see how different threshold values affect the performance of the IPFT. Specifically, we plan to further investigate the resource usage metric threshold for triggering the node replication and task migration and the time intervals for monitoring the time series metrics. 

\section*{Acknowledgment}
This work is part of the ACCORDION and CHARITY projects that have received funding from the European Union’s Horizon 2020 research and innovation programme under grant agreements No 871793" and No 101016509".

\bibliographystyle{unsrt}  
\bibliography{references}

\end{document}